\documentclass[doublecol,reprint]{epl2mod}

\usepackage{epsfig}
\usepackage{amsmath}
\usepackage{amssymb}
\usepackage{mathrsfs}
\usepackage{comment}
\usepackage{color}
\usepackage{braket}
\newcommand{\be}{\begin{equation}}
\newcommand{\ee}{\end{equation}}
\newcommand{\ba}{\begin{aligned}}
\newcommand{\ea}{\end{aligned}}
\renewcommand{\revision}[1]{#1}

\title{New insights into the entanglement of disjoint blocks}
\shorttitle{New Insights into the entanglement of disjoint blocks} 

\author{Maurizio Fagotti\inst{1,2}%\thanks{E-mail: \email{Maurizio.Fagotti@physics.ox.ac.uk}}
}
\shortauthor{M. Fagotti}

\institute{                    
  \inst{1} Dipartimento di Fisica dell'Universit\`a di Pisa and INFN - Pisa 56127, Italy\\
  \inst{2} The Rudolf Peierls Centre for Theoretical Physics, University of Oxford -  Oxford OX1 3NP, United Kingdom
}

\abstract{
We study the entanglement of two disjoint blocks in  
spin-$\frac{1}{2}$ chains obtained by merging solvable models, such as XX and \revision{quantum} Ising models. \revision{We focus on the universal quantities that can be extracted from the R\'enyi entropies $S_\alpha$. The most important information is encoded in some functions denoted by $F_{\alpha}$. We compute $F_2$ and we} show that \revision{$F_\alpha-1$} and \revision{$F_{v.N.}$}, corresponding to the von Neumann entropy, can be \revision{negative}, in contrast to what observed in all models examined so far. An exact relation between the entanglement of disjoint subsystems in the XX model and that in a chain embodying two \revision{quantum} Ising models is a by-product of our investigations.
}

\begin{document}

\maketitle

In the last two decades, the entanglement entropy has become one of the most important indicators of critical behavior in many-body systems~\cite{Amico:2008,Latorre:2009}.
In the neighborhood of a quantum phase transition the area law~\cite{Eisert:2010}, \emph{i.e.}  the proportionality between the entropy and the contact surface with the remainder of the system, does not work in 1D, where the entropy turns out to be proportional to the logarithm of the characteristic subsystem's length. The factor in front of the logarithm is generally universal~\cite{Holzhey:1994,Refael:2004, Igloi:2007} and, in conformal systems, is proportional to the central charge~\cite{Calabrese:2004}. In fact, the entanglement entropy provides one of the most accurate ways of detecting the value of the central charge.

Many recent works~\cite{Caraglio:2008,Furukawa:2009,Calabrese:2009, Fagotti:2010, Igloi:2010, Alba:2010,Calabrese:2011,Alba:2011} have focused on the entanglement of disjoint intervals in conformal systems, because it is also sensitive to universal details of the conformal field theory (CFT) different from the central charge. R\'enyi entropies $S_\alpha=\frac{1}{1-\alpha}\log\mathrm{Tr}\rho^\alpha$ of two disjoint blocks of lengths $\ell_1$ and $\ell_2$ at the distance $r$, as long as $\alpha$ is integer, are proportional to the logarithm of the four-point function of particular twist fields~\cite{Calabrese:2009}.  For any given $\alpha$,  a (bounded) universal function of the four-point ratio $x=\frac{\ell_1 \ell_2}{(\ell_1+r)(\ell_2+r)}$, usually denoted by $F_\alpha(x)$, is the main information, besides the central charge $c$, that can be extracted from the R\'enyi entropy 
\be\label{eq:Salpha}
S_\alpha=\frac{1+\alpha}{\alpha}\frac{c}{6}\log\Bigl|\frac{\ell_1\ell_2 r (\ell_1+r+\ell_2)}{(\ell_1+r)(r+\ell_2)}\Bigr|+\frac{\log F_\alpha(x)}{1-\alpha}+2 c^\prime_\alpha\, .
\ee
The invariance of the entropies under interchanging the subsystem with the rest manifests in the identity \mbox{$F_\alpha(x)=F_\alpha(1-x)$}. The entanglement of  a single spin block can be recovered in the limit as $r$ approaches $\infty$, and hence $x\rightarrow 1$. Indeed, when the blocks are far enough, the reduced density matrix of the two blocks is expected to factorize in the tensor product of the reduced density matrices of each block (interaction is local). Thus, the R\'enyi entropies (in particular the entanglement entropy) become simply the sum of the entropies of the blocks.
This means that, normalizing the universal function in such a way that $F_\alpha(0)=F_\alpha(1)=1$, the constant $c^\prime_\alpha$ in eq.~\eqref{eq:Salpha} is exactly the additive constant of the single interval~\cite{Calabrese:2004}
\be\label{eq:single}
S_\alpha=\frac{1+\alpha}{\alpha}\frac{c}{6}\log \ell+c^\prime_\alpha\, .
\ee
For non-interacting fermions $F_\alpha(x)$ is equal to $1$ for any $x$~\cite{Casini:2009a}, however this is peculiar to the fermionic representation of non-interacting models~\cite{Fagotti:2010}, whilst in the spin representation, as exactly computed in refs.~\cite{Calabrese:2009,Calabrese:2011} in CFT, $F_\alpha(x)$ is a complicated function. It has been determined  only for integral values of $\alpha$ for the free boson compactified on a circle~\cite{Calabrese:2009,Alba:2011} and for the \revision{quantum} Ising model~\cite{Alba:2010,Calabrese:2011}. The universal function $F_{v.N.}(x)\equiv\bigl.\partial_\alpha F_\alpha(x)\bigr|_{\alpha=1}$, corresponding to the von Neumann entropy, is still unknown for any model.
 
One of the goals of this work is to provide exact results for $F_\alpha(x)$, especially for $F_2(x)$, in a family of models whose continuum limit is conformal. In particular, the \mbox{finite-size} scaling can be obtained by substituting any length $\ell$ with the chord length $\frac{L}{\pi}\sin\frac{\pi \ell}{L}$~\cite{Calabrese:2004}, where $L$ is the chain's size. Incidentally, in refs.~\cite{Alcaraz:2011,Ibanez:} it has been shown that the excited states associated to primary operators in a CFT have instead peculiar finite-size scalings.
The results presented in this Letter are the first analytic ones obtained in a framework different from CFT.

In light of the numerical investigations~\cite{Alba:2010,Fagotti:2010,Alba:2011} and of the CFT predictions~\cite{Calabrese:2009,Calabrese:2011}, one could wonder whether some properties, always observed in $F_\alpha(x)$, are indeed general. We are referring to the \emph{positivity} of $F_\alpha(x)-1$ and of $F_{v.N.}(x)$, as well as to the \emph{concavity} of $F_{v.N.}(x)$. The latter could be thought as a consequence of some general property, in the same way as strong subadditivity implies that the von Neumann entropy of a single interval is a concave function of the block's length~\cite{Casini:2004,Fagotti:a}. 
As a matter of fact, some of the models considered here have $F_\alpha(x)\leq 1$, as well as $F_{v.N.}(x)\leq 0$. This is rather surprising
since in ref.~\cite{Calabrese:2011} it has been shown that the first order of the short-length expansion of $F_\alpha(x)$\revision{, corresponding to the limit in which the distance between the blocks is much larger than their lengths,} is positive.
We will see that the two things are in fact compatible.

The key to our results lies in the non-locality of the fermionic mapping of spin-$\frac{1}{2}$ chains.
As discussed in ref.~\cite{Igloi:2010}, and then explained in detail in ref.~\cite{Fagotti:2010}, the entanglement of disjoint subsystems in the fermionic representation of non-interacting spin-$\frac{1}{2}$ chains is different from that in the spin representation, because the Jordan-Wigner transformation
\be\label{eq:J-W}
c_l^\dag=\prod_{j<l}\sigma_j^z \sigma_l^+\, ,\qquad \text{with}\quad \sigma_l^+=\frac{\sigma_l^x+i \sigma_l^y}{2}\, ,
\ee
 is non-local. Igl\'oi and Peschel in ref.~\cite{Igloi:2010} have given a clear evidence of this fact by comparing  the sub-lattice entanglement (\emph{e.g.} the entanglement of the subsystem consisting of all even sites) in the two representations: the entanglement entropy, which is extensive in both cases, has a different slope. In spite of this, the consequences of the inequivalence are often surprising, and the models considered here constitute further examples of counterintuitive behaviors.
 
Finally, studying the entanglement of disjoint subsystems in a chain embodying two \revision{quantum} Ising models, we find an exact correspondence with the XX model, which complements the results obtained by Igl\'oi and Juh\'asz in ref.~\cite{Igloi:2008a} for the entanglement of spin blocks.

%%%%%%%%%%%%
\section{The model}   %
%%%%%%%%%%%%

\revision{We consider a class of spin-$\frac{1}{2}$ chains closely related to the models introduced by Suzuki in ref.~\cite{Suzuki:1971}.
This is an extension of the XY model in which the Hamiltonian has a longer-range interaction, and yet it reduces to free-fermions after the \mbox{J-W transformation~\eqref{eq:J-W}}.  
The strategy is to ``move away'' the spin operators adding a \mbox{$\sigma_z$-string} between those that are non-local in the J-W fermions ($\sigma^{x(y)}$) , 
so that no string appears in the fermionic representation, \emph{e.g.}
\be\label{eq:shift}
\sigma_l^{x}\sigma_{l+1}^{x}\rightarrow \sigma_l^{x}\sigma_{l+1}^z\cdots \sigma_{l+n-1}^z \sigma_{l+n}^{x}\, .
\ee 
We focus on chains that can be constructed by merging XX and Ising models, which we call ``constitutive models''. We consider in particular the Hamiltonians
} 
\be\label{eq:examples}
\begin{cases}
\mathrm{XX}^n&:
\sum_l \sigma_l^+\sigma_{l+1}^z\cdots \sigma_{l+n-1}^z\sigma_{l+n}^-+ h.c.\\
\mathrm{Is}^n&:\sum_l \sigma_l^x\sigma_{l+1}^z\cdots \sigma_{l+n-1}^z\sigma_{l+n}^x+\sigma_l^z\\
\mathrm{XX}\times \mathrm{Is}&:\sum_l \sigma_l^x\sigma_{l+1}^z\sigma_{l+2}^x+ \sigma_{2l-1}^y\sigma_{2l}^z\sigma_{2l+1}^y+ \sigma_{2l}^z\, ,
\end{cases}
\ee
where the names (on the left)  refer to the \revision{constituents: the Hamiltonians are sum of XX and Ising models that act on different fermionic degrees of freedom; the constitutive Hamiltonians are} obtained by restricting the sum \revision{in eq.~\eqref{eq:examples}} over the indices that have the same remainder after the division by \revision{the number} $n$ \revision{of sub-chains}. 
\revision{The models in eq.~\eqref{eq:examples} can be solved by diagonalizing separately the $n$ commuting Hamiltonians.}
For example, \revision{by merging two quantum Ising models we eventually get}
\be\label{eq:Is2}
H_{\mathrm{Is}^2}=\sum_{k}\varepsilon_{k}\Bigl(b^\dag_{k,e} b_{k,e}+b^\dag_{k,o} b_{k,o}-1\Bigr)\, ,
\ee
where $b_{k,e}$ and $b_{k,o}$ are the Bogolioubov fermions that diagonalize the Ising model, constructed with the even and \revision{the} odd J-W fermions, respectively; $\varepsilon_k\equiv\varepsilon_{k,e}=\varepsilon_{k,o}$ is the dispersion relation. 
By looking at the low-energy excitations, we find that \revision{the dispersion relation is approximately linear and} the number of chiral modes \revision{(\emph{i.e.} the number of zeros of the dispersion relation above the ground state)} is equal to the sum of the chiral modes of the constitutive models.
This observation is crucial to infer the value of the central charge of the underlying CFT, which turns out to be the sum of the central charges of the $n$ constitutive models. Thus, the long distance physics in the examples~\eqref{eq:examples} is expected to be described by a CFT with central charge $n$, $\frac{n}{2}$, and $\frac{3}{2}$, respectively ($c=1$ in the XX model and $c=\frac{1}{2}$ in the Ising one). 

We point out that all Hamiltonians with the same structure as in \revision{eq.~}\eqref{eq:examples} are equivalent  (up to involution operators multiplying the boundary terms) to the corresponding Hamiltonians without the $\sigma_z$-string in eq.~\eqref{eq:shift}, \emph{e.g.}
\be\label{eq:2copyIs}
H_{\mathrm{Is}^2}\sim H_{\mathrm{Is}}\oplus H_{\mathrm {Is}}=\sum_l \sigma_l^x\sigma_{l+2}^x+\sigma_l^z\, .
\ee
To avoid confusion we call the latter ones ``$n$-copy Hamiltonians", although in general they could be sum of different Hamiltonians.
The equivalence is realized by the transformation
\be\label{eq:equivalence}
\sigma_{n l +s}^+\rightarrow \prod_{i<nl+s}\sigma_i^z\prod_{s^\prime<s}\bigl(\prod_j \sigma_{n j+s^\prime}^z\bigr)\prod_{j<l}\sigma_{n j+s}^z\sigma_{n l+s}^+\, .
\ee
\revision{This can be seen as follows:
starting from the $n$-copy Hamiltonian, we apply $n$ J-W transformations~\eqref{eq:J-W} restricted to each sub-chain (the last product in eq.~\eqref{eq:equivalence}). The new Hamiltonian appears like the fermionic representation of the merged one, but the operators, which behave like fermions inside of their own sub-chain, commute with those of the other sub-chains. 
We indicate with $a^z_{[s]}$ the complete \mbox{$\sigma_z$-string} of the sub-chain $s$ ($a^z_{[s]}$ commutes with $H$ and anticommutes with the fermions in $s$). In order to restore the correct algebra, we multiply the operators in $s$ by $a^z_{[s^\prime]}$ if and only if the operators in $s^\prime$ are not multiplied by $a^z_{[s]}$ (this is the origin of the central product in eq.~\eqref{eq:equivalence}). Finally, the inverse J-W transformation (the first product in eq.~\eqref{eq:equivalence}) maps the Hamiltonian into the merged one.}

The model obtained by merging $n$ constitutive chains and the corresponding $n$-copy model have different entanglement features.
We sum up firstly the properties of the $n$-copy model.

%%%%%%%%%%%%%%
\section{$n$-copy model}%
%%%%%%%%%%%%%%

Because we are considering spatial entanglement,  the entanglement entropy in the ground state of an $n$-copy Hamiltonian is sensitive to its simple structure: the reduced density matrix of any subsystem is the tensor product of the reduced density matrices restricted to the subspaces in which the constitutive Hamiltonians act, \emph{e.g.} in the model $H_{\mathrm{Is}}\oplus H_{\mathrm{Is}}$, eq.~\eqref{eq:2copyIs}, we find the factorization
\be\label{eq:shuffledfact}
\rho_A^{\mathrm{Is}\oplus \mathrm{Is}}=\rho_{A_e}^{\mathrm Is}\otimes \rho_{A_o}^{\mathrm{Is}}\, ,
\ee
where $A_e$ ($A_o$) is the subsystem consisting of the even (odd) sites of $A$, and $\rho^{\mathrm{Is}}$ is the reduced density matrix (RDM)  in the Ising model. In particular this means that
\be\label{eq:shuffled}
\ba
c=\sum_i c_{(i)}&&c^\prime_{\alpha}=\sum_i\Bigl((c^\prime_\alpha)_{(i)}-\frac{1+\alpha}{\alpha}\frac{c_{(i)}}{6} \log n\Bigr) 
\ea
\ee
and
\be
F_\alpha(x)=\prod_i F^{(i)}_\alpha(x)\, ,
\ee
where the index $i$ runs over the constitutive models  (\emph{cf.} eq.~\eqref{eq:Salpha}). The underlying CFT is the tensor product of the theories describing the low-energy excitations of the constitutive models.
No auxiliary information can be extracted from the entanglement in the ground state of $n$-copy Hamiltonians, apart from those already known from the constitutive models.

%%%%%%%%%%%%%%%%
\section{Two disjoint blocks} %
%%%%%%%%%%%%%%%%

We now come back to our model. Because the entanglement of a single interval is equal both for spins and J-W fermions (the J-W transformation, although non local, mixes only degrees of freedom inside of the block), eq.~\eqref{eq:shuffled} continues to be valid: we do not find new information studying the block entanglement. 
However, in chains with periodic boundary conditions, the RDM of disjoint blocks is different from the corresponding density matrix in the fermionic representation. In ref.~\cite{Fagotti:2010} it has been shown that the RDM of two disjoint blocks $A\equiv A_1\cup A_2$ (we call $B_1$ and $B_2$ the blocks between $A_1$ and $A_2$) is equivalent to the sum of four operators (two genuine density matrices and two ``fake" ones) 
\be\label{eq:rho2blocks}
\rho_{A}=\frac{\rho_{A}^{fer}+a_{A_2}^z \rho_{A}^{fer}a_{A_2}^z}{2}+\braket{a^z_{B_1}}\frac{\rho_A^{(B_1)}-a_{A_2}^z\rho_A^{(B_1)}a_{A_2}^z}{2}\, ,
\ee
which for non-interacting models are exponentials of quadratic forms. The first operator is the fermionic density matrix. We indicated with $a^z_R$ the $\sigma_z$-string, product of every $\sigma_z$ in $R$. The fake density matrix is defined as
\be
\braket{a^z_{B_1}}\rho^{(B_1)}\equiv \mathrm{Tr}_{B_1\cup B_2}[\ket{\Psi_0}\bra{\Psi_0}a^z_{B_1}]\, ,
\ee
where $\ket{\Psi_0}$ is the ground state. In fact, the main difference from the $n$-copy model is that the factorization, shown in the example~\eqref{eq:shuffledfact}, holds now for each operator in eq.~\eqref{eq:rho2blocks} separately (in particular the expectation value of the string factorizes).
The second moment $\mathrm{Tr}\rho^2_A$ in a chain constructed with $n$ identical constitutive models is given by (see ref.~\cite{Fagotti:2010} for the basic case $n=1$)
\be\label{eq:nconst}
\mathrm{Tr}\rho_A^2=\frac{\{\Gamma_1^2\}^n+\{\Gamma_1,\Gamma_2\}^n}{2}+\braket{a^z_{B_1}}^{2 n}\frac{\{\Gamma_3^2\}^n-\{\Gamma_3,\Gamma_4\}^n}{2}\, ,
\ee
where we used the same notation as in ref.~\cite{Fagotti:2010} for the correlation matrices: $\Gamma_1\equiv\mathrm I-\braket{a\otimes a}$ is the fermionic correlation matrix, with $a_i$ the Majorana fermions defined as $a_{2i-1}=c^\dag_i+c_i$ and $a_{2i}=i(c_i-c^\dag_i)$;
$\Gamma_2=P_2\Gamma_1 P_2$  and $\Gamma_4=P_2\Gamma_3 P_2$, where $P_2$ inverts the sign of the components corresponding to the second block $A_2$; 
$\Gamma_3$ is given by 
\be
\Gamma_3=\Gamma_1-\Gamma_{A B_1}\Gamma_{B_1 B_1}^{-1}\Gamma_{B_1 A}\, ,
\ee
where the double subscripts take into account restrictions to rectangular correlation matrices, \emph{i.e.} the first (second) subscript identifies the region where the row (column) index runs  (\emph{e.g.} $\Gamma_{A A}\equiv \Gamma_1$). $\{\Gamma,\Gamma^\prime\}$  is the product of the eigenvalues of $(\mathrm I+\Gamma\Gamma^\prime)/2$ with halved degeneracy
(see ref.~\cite{Fagotti:2010} for further details).
We stress that in the correlation matrices above each length is divided by $n$, indeed the fermionic space of the constitutive models is the $n^{th}$ root of the total space.
The generalization to distinct constitutive models is straightforward: we must substitute powers with products in eq.~\eqref{eq:nconst}. 

In order to get the universal function $F_2(x)$, because $\{\Gamma_1^2\}\equiv\{\Gamma_1,\Gamma_1\}$ is the fermionic analogous of eq.~\eqref{eq:nconst}, corresponding to $F_2(x)=1$ (free fermions), we can divide $\mathrm{Tr}\rho_A^2$ by $\{\Gamma_1^2\}^n$, which takes into account the actual value   
of the central charge $c$ and of the constant $c^\prime_2$ (\emph{cf.} eq.~\eqref{eq:Salpha} and eq.~\eqref{eq:shuffled})
\be\label{eq:F2n}
F_2^{(n)}(x)=\frac{1}{2}+\frac{\{\Gamma_1,\Gamma_2\}^n+\braket{a^z_{B_1}}^{2 n}(\{\Gamma_3^2\}^n-\{\Gamma_3,\Gamma_4\}^n)}{2 \{\Gamma_1^2\}^n}\, .
\ee
It is remarkable that eq.~\eqref{eq:F2n} gives direct access to the universal function, which is a subleading term of the R\'enyi entropy $S_2$ (\emph{cf.} eq.~\eqref{eq:Salpha}).
We point out that the knowledge of $F_2^{(n)}(x)$ for $n=1,2,3$ is sufficient to get   the function for any $n$.
However, if we are interested in  $F_3^{(n)}(x)$ (or in $F_4^{(n)}(x)$) we need 4 (or 18) values of $n$ to compute the universal function for any $n$, because of the increasing number of terms in the formulae analogous to eq.~\eqref{eq:nconst} (see eqs. (58,59) of ref.~\cite{Fagotti:2010}).

Equations~\eqref{eq:rho2blocks}, \eqref{eq:nconst}, and \eqref{eq:F2n} are general: they hold for any set of non-interacting constitutive models. Equation~\eqref{eq:rho2blocks} is valid also in the presence of  interaction. We specialize now the formulae to XX and Ising \revision{sub-chains}.

%%%%%%%%%%%
\section{XX model}%
%%%%%%%%%%%

The entanglement entropy of two disjoint blocks in the XX model has been investigated in relatively small chains by Furukawa {\etal} in ref.~\cite{Furukawa:2009} by exact diagonalization techniques. In ref.~\cite{Fagotti:2010}  the CFT prediction~\cite{Furukawa:2009,Calabrese:2009}
\be
F_2^{XX}(x)=\frac{1+\sqrt{1-x}+\sqrt{x}}{2}
\ee
has been checked numerically in the thermodynamic limit. Now, in order to find the analogous results for $n>1$, \emph{i.e.} for the ground state of the Hamiltonian
\be\label{eq:XXn}
H_{\mathrm{XX}^n}\equiv\sum_l \sigma_l^+\sigma_{l+1}^z\cdots \sigma_{l+n-1}^z\sigma_{l+n}^-+ h.c.\, ,
\ee
we should determine the behavior of each of the terms in eq.~\eqref{eq:nconst}. The correlation matrix $\Gamma_1$ is a Toeplitz matrix in which a block of rows (associated to the region $B_1$) and the corresponding block of columns have been removed. 
Therefore, the basic problem of computing the R\'enyi entropies of disjoint blocks of the same length in the fermionic representation of XX chains can be traced back to the asymptotic behavior of the determinant of large block-Toeplitz matrices with Fisher-Hartwig singularities. However, we are not aware of useful mathematical theorems or conjectured for this kind of matrices. In addition, the other matrices appearing in eq.~\eqref{eq:nconst} have structures even more complicated.
\begin{figure}
\onefigure[width=0.44\textwidth]{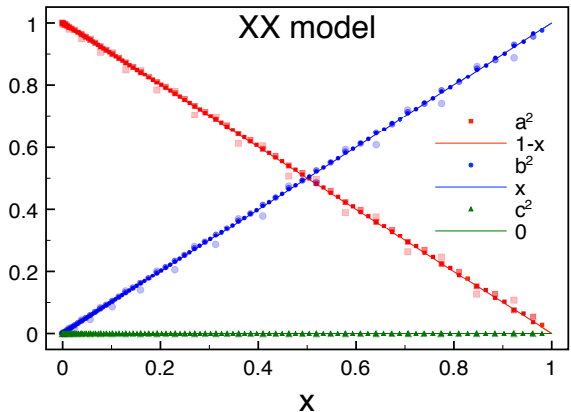}
\caption{The terms in eq.~\eqref{eq:XXprop} for $\ell+r=25,50,100$; symbols are sorted by increasing opacity and decreasing size. In the legend $a=\{\Gamma_1,\Gamma_2\}/\{\Gamma_1^2\}$, $b=\braket{a^z_{B_1}}^{2}\{\Gamma_3^2\}/\{\Gamma_1^2\}$, and $c=\braket{a^z_{B_1}}^{2}\{\Gamma_3,\Gamma_4\}/\{\Gamma_1^2\}$. Data converge quickly to the functions proposed in eq.~\eqref{eq:XXprop} (solid lines).
}
\label{fig:scalingXX}
\end{figure}

In any case, we provide numerical evidence that each of the terms summed in eq.~\eqref{eq:F2n} is a simple function of $x$, namely
\be\label{eq:XXprop}
\ba
\frac{\{\Gamma_1,\Gamma_2\}}{\{\Gamma_1^2\}}&\sim\sqrt{1-x}\qquad
\frac{\braket{a^z_{B_1}}^{2}\{\Gamma_3^2\}}{\{\Gamma_1^2\}}\sim\sqrt{x}\, ,\\
\ea
\ee
while the last term of eq.~\eqref{eq:F2n} is subleading.
In fig.~\ref{fig:scalingXX} the proposed scaling functions are checked against numerics \revision{in the thermodynamic limit} for various subsystem's configurations with blocks of the same length $\ell_1=\ell_2=\ell$. The agreement is excellent. 
In general, by inserting  eq.~\eqref{eq:XXprop} into eq.~\eqref{eq:F2n} we get
\be\label{eq:Fn_2}
F_2^{(n)}(x)=\frac{1+(1-x)^\frac{n}{2}+x^{\frac{n}{2}}}{2}\, .
\ee
For $n=2$ the function simplifies to $F_2^{(2)}(x)=1$.
\begin{figure}
\onefigure[width=0.44\textwidth]{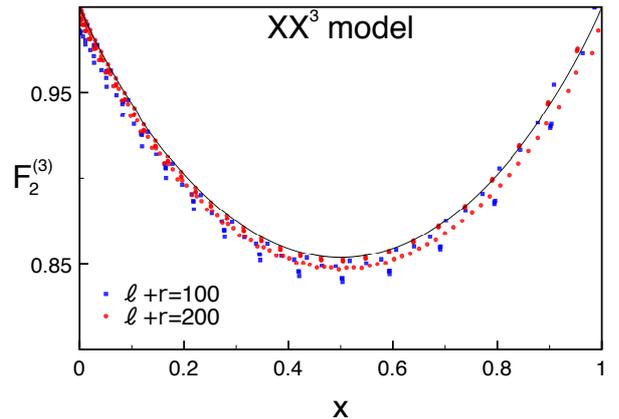}
\caption{The universal function $F_2(x)$ for the $\mathrm{XX}^3$ model (eq.~\eqref{eq:XXn} with $n=3$). The universal function is evidently smaller than 1. The continuous curve is eq.~\eqref{eq:Fn_2} with $n=3$. }
\label{fig:F3_2}
\end{figure}
However, for $n>2$ the universal function becomes \emph{smaller than $1$}, and in the limit $n\rightarrow\infty$ we get $F_2^{(\infty)}(x)=\frac{1}{2}$ for any $x\in(0,1)$. In fig.~\ref{fig:F3_2} we report the universal function corresponding to $n=3$ for two values of $\ell+r$. 

Equations~(\ref{eq:F2n},\ref{eq:Fn_2}) are the main results of this Letter.
In particular eq.~\eqref{eq:Fn_2} is the first expression for the universal function $F_2(x)$ obtained in a framework different from CFT. Furthermore, for $n>2$ it is the first example in which $F_2(x)<1$. 

Calabrese{\etal} in ref.~\cite{Calabrese:2011} have determined the short-length expansion for the free boson and Ising model 
\be\label{eq:expansionC}
F_{2}(x)=1+\mathcal N\Bigl(\frac{x}{16}\Bigr)^y+O(x)+\dots\, ,
\ee
where $y$ is the scaling dimension of an operator coming from the operator product expansion of twist fields, and $\mathcal N$ is the number of inequivalent correlation functions given the same contribution. 
By series expanding eq.~\eqref{eq:Fn_2} for small $x$ we find
\be\label{eq:expansionM}
F_2^{(n)}(x)=1+\frac{1}{2}x^{\frac{n}{2}}-\frac{n}{4}x+\dots\, .
\ee
As observed in ref.~\cite{Calabrese:2011}, when there is no operator in the theory whose correlation function can contribute to $O(x)$, the linear term in eq.~\eqref{eq:expansionC} must cancel  with the $O(x)$ term coming from the function multiplying $F_\alpha(x)$ in $\mathrm{Tr}\rho^\alpha$, \emph{i.e.} it must be equal to $-\frac{1+\alpha}{\alpha} \frac{c}{6}x$. This is exactly what happens here, being the central charge $c=n$ (\emph{cf.} eq.~\eqref{eq:shuffled}). If eq.~\eqref{eq:expansionC} can be applied, by direct comparison between eq.~\eqref{eq:expansionM} and eq.~\eqref{eq:expansionC} we recognize $y=\frac{n}{2}$ and $\mathcal N=4^n/2$. 
The fact that $F^{(n)}_2(x)\leq 1$ for $n>2$ is essentially due to the $O(x)$ contribution in eq.~\eqref{eq:expansionM}. 

Incidentally, the short-length expansion in the $n$-copy model is characterized by the same exponent $y=\frac{1}{2}$ of the XX model  and $\mathcal N=2 n$, being the $n$-copy model sum of $n$ identical models.

From the analytic continuation of the short-length expansion of $F_{\alpha}(x)$, it can be deduced the expansion of $F_{v.N.}(x)$~\cite{Calabrese:2011}
\be
F_{v.N.}(x)=\mathcal N\Bigl(\frac{x}{4}\Bigr)^y\frac{\sqrt{\pi}\Gamma(y+1)}{4\Gamma(y+\frac{3}{2})}-\frac{c}{3}x +\dots\, .
\ee
By substituting the value of $y$ we get
\be
F_{v.N.}^{(n)}(x)=(4 x)^{\frac{n}{2}}\frac{\sqrt{\pi}\Gamma(\frac{n+2}{2})}{8 \Gamma(\frac{n+3}{2})}-\frac{c}{3}x+\dots\, .
\ee
In particular the universal function $F_{v.N.}(x)$ vanishes for $n=2$. 
For $n>2$ $F_{v.N.}^{(n)}(x)$ \emph{is negative} (at least for small enough $x$), in contrast to what observed up to this \mbox{time~\cite{Furukawa:2009,Alba:2009,Alba:2010}}. We point out that if $F_{v.N.}(x)$ is negative then the first order of the small-length expansion must be linear. This is compatible with the constraints given by strong subadditivity~\cite{Fagotti:a}.

The models obtained by merging XX chains, eq.~\eqref{eq:XXn}, have a further property: they commute for different values of $n$, and in particular they commute with the XX Hamiltonian ($n=1$). 
Thus, from eq.~\eqref{eq:Fn_2} and eq.~\eqref{eq:shuffled} with $c_{(i)}=1$ and $(c^\prime_{2})_{(i)}=c^\prime_2$ the additive constant for the block R\'enyi entropy $S_2$ (see ref.~\cite{Jin:2004} for the analytic expression), we get the entropy of two disjoint intervals in some excited states of the XX model. In this way we have extended the analysis of refs.~\cite{Alba:2009,Calabrese:2011b,Calabrese:2011c}, focused on the entanglement of spin blocks.

%%%%%%%%%%%%
\section{Ising model} % 
%%%%%%%%%%%%

We now consider the merging of $n$ Ising chains
\be\label{eq:Isn}
H_{\mathrm{Is}^n}\equiv\sum_l\sigma_l^x\sigma_{l+1}^z\cdots\sigma_{l+n-1}^z\sigma_{l+n}^x+\sum_l\sigma_l^z\, .
\ee
The first approximate numerical results for the entanglement entropy of two disjoint blocks in the Ising model ($n=1$) have been obtained for $\alpha=2$ in ref.~\cite{Alba:2010}
by using a tree 
tensor network algorithm and Monte Carlo simulations.
\begin{figure}
\onefigure[width=0.44\textwidth]{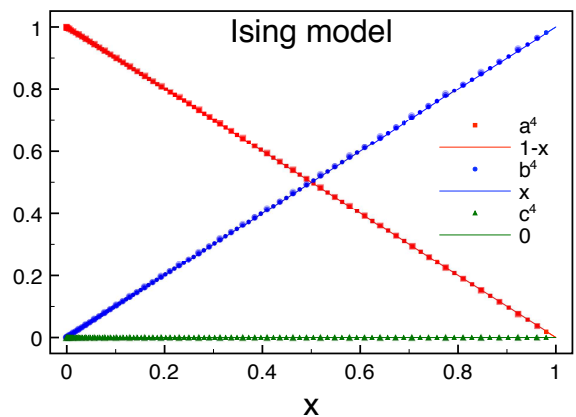}
\caption{The terms in eq.~\eqref{eq:Isprop} for $\ell+r=25,50,100$; symbols are sorted by increasing opacity and decreasing size. $a$, $b$, and $c$ are the same quantities defined in fig.~\ref{fig:scalingXX}. Data converge quickly to the functions proposed in eq.~\eqref{eq:Isprop} (solid lines). }
\label{fig:scalingIs}
\end{figure}
In ref.~\cite{Fagotti:2010} the CFT prediction~\cite{Alba:2010,Calabrese:2011}
\be
F_2^{Is}(x)=\frac{1+\sqrt[4]{1-x}+\sqrt[4]{x}}{2}
\ee
has been checked against exact numerical data. As done before for the XX model, in order to get the universal function $F_2^{(n)}(x)$ we must determine the behavior of each of the terms in eq.~\eqref{eq:F2n}. The result, as shown in fig.~\ref{fig:scalingIs}, has the same structure as in the XX model (\emph{cf.} eq.~\eqref{eq:XXprop})
\be\label{eq:Isprop}
\ba
\frac{\{\Gamma_1,\Gamma_2\}}{\{\Gamma_1^2\}}&\sim\sqrt[4]{1-x}\qquad
\frac{\braket{a^z_{B_1}}^{2 n}\{\Gamma_3^2\}^n}{\{\Gamma_1^2\}^n}\sim\sqrt[4]{x}\, ,\\
\ea
\ee
and the last term of eq.~\eqref{eq:F2n} is again subleading.
The universal function corresponding to merging $n$ Ising models is then given by
\be\label{eq:Fn_2Is}
F_2^{(n)}(x)=\frac{1+(1-x)^\frac{n}{4}+x^{\frac{n}{4}}}{2}\, .
\ee
This formula is identical to eq.~\eqref{eq:Fn_2}, provided that $n$ is substituted by $\frac{n}{2}$. In particular the central charge $c$ and the exponent $y$ are both equal to $\frac{n}{2}$. For $n=2$, namely
\be\label{eq:Is2}
H_{\mathrm{Is}^2}=\sum_{l}\sigma_l^x\sigma_{l+1}^z\sigma_{l+2}^x+\sum_l \sigma_l^z\, ,
\ee
 both the additive constants $c^\prime_\alpha$ of the single interval (\emph{cf.} eq.~\eqref{eq:single}) and the universal function $F_2(x)$ are equal to those corresponding to the XX model. 
\begin{figure}
\onefigure[width=0.44\textwidth]{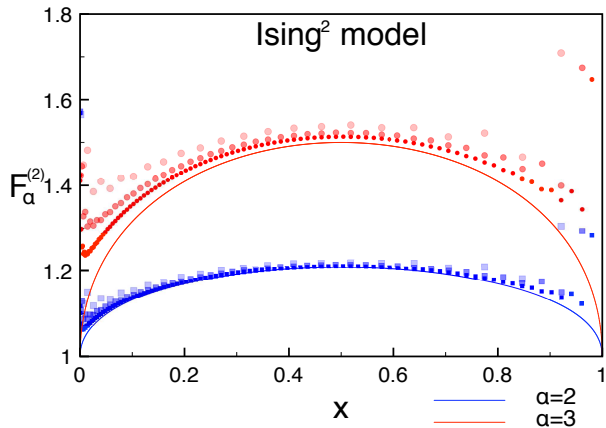}
\caption{The universal functions $F_2^{(2)}(x)$ and $F_3^{(2)}(x)$ for $\ell+r=25,50,100$ in the Ising$^2$ chain (eq.~\eqref{eq:Isn} with $n=2$); symbols are sorted by increasing opacity and decreasing size. Data converge to the universal functions of the XX model (continuous curves).
}
\label{fig:FIs2}
\end{figure}
In fig.~\ref{fig:FIs2} we report both $F_2^{(2)}(x)$ and $F_3^{(2)}(x)$, showing agreement with the CFT predictions relative to the XX model.
We expect this remains true for any $\alpha$ \revision{(\emph{cf.} eq.~\eqref{eq:Salpha})}. 
The relation between the entanglement in the XY model and that in the Ising model has been already discussed in ref.~\cite{Igloi:2008a}, where the authors focused on the entanglement entropies of spin blocks.
Here we have established a correspondence between the entanglement of disjoint blocks in the XX model and that in two merged Ising chains, eq.~\eqref{eq:Is2}. The two models display different corrections to the scaling, however the asymptotic behavior of R\'enyi entropies coincides.

In contrast to the XX model, the Hamiltonians~\eqref{eq:Isn} do not commute with the Ising Hamiltonian.

The third example in eq.~\eqref{eq:examples} shows the merging of an XX chain with an Ising one. The Hamiltonian is not translational invariant, however the symmetry will be eventually recovered in the continuum limit, as it happens for the $2$-copy Hamiltonian. Equation~\eqref{eq:F2n} still holds for this kind of models and in general we find
\be
F_2^{(n_{\mathrm{XX}},n_{\mathrm{Is}})}(x)=\frac{1+(1-x)^{\frac{2n_{\mathrm{XX}}+n_{\mathrm{Is}}}{4}}+x^{\frac{2n_{\mathrm{XX}}+n_{\mathrm{Is}}}{4}}}{2}\, ,
\ee
where $n_{\mathrm{XX}}$ and $n_{\mathrm{Is}}$ are the numbers of XX and Ising constitutive models, respectively. 
Because of the correspondence between XX and merged Ising chains, this result does not add significant information.
However, in general, merging different chains results in universal functions that are not trivially related to the constitutive ones.  

%%%%%%%%%%%%%%%%%%
\section{Summary and discussion}%
%%%%%%%%%%%%%%%%%%

We have investigated the entanglement entropy of two disjoint blocks in exactly solvable spin-$\frac{1}{2}$ chains. 
We derived the universal function $F_2(x)$ for the XX and the Ising model, reproducing the known results from CFT. In addition, we computed $F_2(x)$ for an entire class of chains that embody models with a free-fermion representation. We have shown that $F_\alpha(x)$ can be smaller than $1$, as well as $F_{v.N.}(x)$ negative, by providing examples in which this indeed happens.
Finally, we revealed a correspondence between the XX model and two merged Ising chains, extending the analysis of ref.~\cite{Igloi:2008a} to the entanglement of disjoint subsystems.

In this paper we considered $\alpha=2$, however
the analysis of R\'enyi entropies with $\alpha>2$ could be useful to understand the connection between the CFT formalism and the method  of ref.~\cite{Fagotti:2010}, based on free-fermion techniques, perhaps making progress toward the problem of computing the universal function $F_{v.N.}(x)$ of the von Neumann entropy.

In light of the existence of models in which $F_{v.N.}(x)<0$, it becomes worth determining the lower bound of the universal function, for example investigating the consequences of strong subadditivity \cite{Fagotti:a}. 

Finally, we have seen that the XX Hamiltonian commutes with the models embodying XX chains. However, it does not commute with the corresponding  $n$-copy Hamiltonians. Thus, it could be interesting to study the features of the stationary state that describes the system late times after the quench resulting from joining together the $n$ decoupled XX \revision{sub-chains} to form an XX chain.

\acknowledgments
I thank Pasquale Calabrese for useful comments and Fabian Essler for discussions. \revision{I thank Jacques Perk for drawing my attention to ref.~\cite{Suzuki:1971}.}

\end{document}